\documentclass{article}
\usepackage{amsmath}
\usepackage{graphics}
\usepackage{amsmath,amssymb,mathtools}
\usepackage{graphics,color,array,calc,rotating,epsfig,psfrag}
\numberwithin{equation}{section}
\usepackage[super,compress]{cite}
\usepackage{bm}
\usepackage{dcolumn}
\usepackage{float}
\usepackage{subfigure}
\usepackage{url}
\usepackage{subfloat}
\usepackage[super,compress]{cite}
\usepackage{hyperref}
\hypersetup{colorlinks,urlcolor=black,citecolor=black,linkcolor=black,filecolor=black}
\usepackage{breakurl}
\usepackage{amsfonts}
\usepackage[mathscr]{eucal}
\def\be{\begin{equation}} \def\ee{\end{equation}}
\def\bea{\begin{eqnarray}} \def\eea{\end{eqnarray}}

\def\bib{B\kern-.05em{I}\kern-.025em{B}\kern-.08em}
\def\btex{B\kern-.05em{I}\kern-.025em{B}\kern-.08em\TeX}

\begin{document}

\begin{center}

{\textbf{\large Neutrino spin oscillations in
 gravitational fields in higher dimensions}}\\

{\small S.A. Alavi$^{\dag,\ddag}$ ,T. Fallahi.Serish$^{\ddag}$\\
$^{\ddag}$\small \emph{Department of physics, Hakim Sabzevari University, P.O. Box 397}\\
\small \emph{Sabzevar, Iran}\\
$^{\dag}$\small \emph{s.alavi@hsu.ac.ir; alaviag.gmail.com}}

\end{center}

\small Neutrino physics is one of the most active fields of research with important implications for particle physics, cosmology and astrophysics. On the other hand, motivated by some theories including string theory, formulation of physical theories in more than four space-time dimensions has been the subject of increasing attention in recent years. Interaction of neutrinos with gravitational fields is one of the interesting phenomena which can lead to transition between different helicity states(spin oscillations). We study neutrino spin oscillations in Schwarzschild and RN backgrounds in higher-dimensional gravitational fields. We calculate the transition probability as a function of time and also study the dependence of the oscillation frequency on the orbital radius. The results help us to
better understand the behavior of gravity and neutrinos in higher dimensions.
\\


\section{\large{Introduction}}

Neutrino physics is one of the most interesting fields of study in various branches of physics from particle physics to cosmology and astrophysics. Since 1930, when Pauli first hypothesized the existence of a neutral particle later called neutrino, much research was done on various aspects of this fundamental particle, leading to the discovery of neutrinos and other developments in subsequent years. The types of neutrinos include electron neutrinos $\nu_{e}$, tau neutrinos $\nu_{\tau}$, muon neutrinos $\nu_{\mu}$, and their antiparticles. Significant differences in the experimental values of neutrino fluxes produced in reactors, accelerators and solar sources, compared to the values predicted by Standard Models, led to several hypotheses, among which the theory of neutrino oscillation was considered more seriously. There are currently three possibilities for neutrino oscillations: in the first type of oscillation, different flavors of neutrinos can be converted to each other, which is important in the physics of solar and atmospheric neutrinos. The second type is the oscillation or transition
between neutrino helicity states of the same flavor, called neutrino spin oscillations. The third one, which is a combination of the previous two types, is called spin-flavor oscillations. The interaction of neutrinos with gravitational fields is one of the interesting phenomena that can lead to transitions between different helicity states (spin oscillations). It was realized more than 20 years ago that external fields drastically change the process of vacuum neutrino oscillations. The influence of the gravitational interaction on neutrino oscillations has been studied in many publications, see Refs.~\citen{l1,l2,l3,l4}. On the other hand, according to some theories such as string theory, the formulation of physical theories in more than four dimensions (4d) of space-time received increasing attention in recent years. Since Einstein proposed his general theory of relativity in 1915, a lot of research has been devoted to unify general relativity (GR) and electromagnetism as two fundamental interactions in nature. However, the early proposals date back to the 1920s, through Kaluza–Klein
theory to unify these interactions \cite{l5,l6} that was a classical unified field theory built in five-dimensional space-time. Recently, motivated by string theory as a requirement for describing a consistent theory of quantum gravity, extra dimensions have been the subject of much attention. Besides string theory, there are some other theories proposing the necessity of extra dimensions.\cite{l7,l8} In this work, we study neutrino spin oscillations in Schwarzschild and Reissner–Nordstrom (RN) backgrounds in higher dimensions. It is worth mentioning that
gravitational measurements in higher dimensions are studied in Ref. \citen{l9}.

\section{\large{Neutrino Spin Oscillations in Gravitational
 Fields}}\label{s1}

\textrm{The relation between the vierbein four-velocity} $ u^{a}=(u^{0},u_{1},u_{2},u_{3}) $ \textrm{ and four-velocity of the particle in its geodesic path}, $U^{\mu}$ is \cite{l10,l11,l12,l13}

\begin{equation}\label{2}
{u^{a}}=e^{a}_{\mu}U^{\mu},
\end{equation}

where $ e^{a}_{\mu} $ are nonzero vierbein vectors and the components of $U^{\mu}$in spherical coordinates are as follows:

\begin{equation}\label{3}
{U^{a}}=(U^{0},U_{r},U_{\theta},U_{\varphi}), U^{0}=\frac{dt}{d\tau}, U_{r}=\frac{dr}{d\tau}, U_{\theta}=\frac{d\theta}{d\tau}, U_{\varphi}=\frac{d\varphi}{d\tau}.
\end{equation}

\textrm{In the relevant metric}, $ U^{\mu} $ \textrm{ is related to the world velocity of the particle through} $ \overrightarrow{U}=\gamma\overrightarrow{v} $  where $ \gamma=\frac{dt}{d\tau} $ and $\tau$ is the proper time. To study the spin oscillation of a particle in a gravitational field we calculate $ G_{ab}=(\overrightarrow{E},\overrightarrow{B}) $ which is the analog tensor of the electromagnetic field and  is defined as

\begin{equation}\label{6}
G_{ab}=e_{a\mu;\nu}e^{\mu}_{b}U^{\nu}, G_{0i}=E_{i}, G_{ij}=-\varepsilon_{ijk}B_{k}.
\end{equation}

$ e_{a\mu;\nu} $ are the covariant derivatives of vierbein vectors:

\begin{equation}\label{7}
e_{a\mu;\nu}=\frac{\partial e_{a\mu}}{\partial_{x^{\nu}}}-\Gamma^{\lambda}_{\mu\nu}e_{a\lambda},
\end{equation}

where $ \Gamma^{\lambda}_{\mu\nu} $ is  the  Christopher symbol. Geodesic equation of a particle in a gravitational field is\cite{l14}

\begin{equation}\label{8}
\frac{d^{2}x^{\mu}}{dq^{2}}+\Gamma^{\mu}_{\sigma\upsilon}\frac{dx^{\sigma}}{dq}\frac{dx^{\upsilon}}{dq}=0,
\end{equation}

where the variable $ q $ parameterizes the particle’s world line. Neutrino spin oscillations frequency is given by the expression $ \overrightarrow{\Omega}=\frac{\overrightarrow{G}}{\gamma} $, where vector $ \overrightarrow{G}$ is defined in the following way\cite{l10,l11,l15}:

\begin{equation}\label{9}
\overrightarrow{G}=\frac{1}{2}(\overrightarrow{B}+\frac{1}{1+u^{0}}[\overrightarrow{E}\times\overrightarrow{u}]).
\end{equation}

\section{\large{Neutrino Spin Oscillations in Schwarzschild Metric in Higher Dimensions}}

Amongst the various types of black hole solutions of Einstein field equations, a natural higher-dimensional generalization of the Schwarzschild metric, also known as the Schwarzschild-Tangherlini metric \cite{l8}, has been assumed to be stable, like its four-dimensional counterpart. The spacetime around such a stationary, neutral, non-rotating and spherically symmetric black hole in $(d + 1)$ dimensions is described by the most general standard form of the spacetime element:

\begin{equation}\label{10}
ds^{2}=A^{2}(r)dt^{2}-A^{-2}(r)dr^{2}-r^{2}d\Omega^{2}_{d-1},
\end{equation}

where $ d\Omega^{2}_{d-1} $ denotes the element of unit $ (d-1)$ -sphere with area $ S_{d-1} $ and we have
\begin{equation}\label{11}
S_{d-1}=\frac{2\pi^{\frac{d}{2}}}{\Gamma[\frac{d}{2}]}, A(r)=\sqrt{1-\frac{\mu_{0}}{r^{d-2}}}.
\end{equation}

The constant parameter $ \mu_{0} $ is related to the mass of the black hole by the following relation\cite{l16}:

\begin{equation}\label{12}
M=\frac{(d-1)S_{d-1}\mu_{0}}{16\pi G_{d+1}},
\end{equation}

where $ G_{d+1}=G_{4}L^{d-3} $ is $(d+1)$-dimensional gravitational constant and L is the size of
the extra dimensions, so

\begin{equation}\label{13}
A(r)=\sqrt{1-\frac{8G_{4}ML^{d-3}\pi^{1-\frac{d}{2}}\Gamma[\frac{d}{2}]}{(d-1)r^{d-2}}}.
\end{equation}

The mass $ M $ is related to  the Schwarzschild  radius through  $ M=\frac{r_{g}}{2} $. It should be mentioned that if a gravitational radius of a black hole is much smaller than the characteristic length of the extra dimensions, then the black hole can be very well described by asymptotically flat solutions\cite{l8,l17,l18,l19}(see also references therein).
For later convenience, we set $ G_{4}=1 $ and define dimensionless variables  $ x=\frac{r}{\ell_{p}},\beta=\frac{r_{g}}{\ell_{p}}$ and $\alpha=\frac{L}{\ell_{p}}$ where $ \ell_{p} $ is the Planck length.

\begin{figure}[H]
\centering
\includegraphics[width=3.5in]{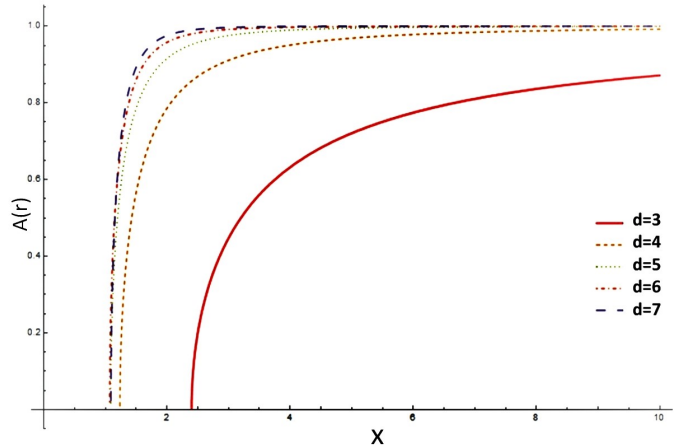}
\caption{\scriptsize (Color online) The $A(r)$ component of the Schwarzschild metric vs. x for different dimensions $ d=3-7 $ and with certain values of $ \beta=2.4 $ and $ \alpha=1.5 $.\label{Fig0}}
\end{figure}

We have plotted the $A(r)$ component of the higher dimensional Schwarzschild metric as a function of $x$ for different spatial dimensions in  Fig. \ref{Fig0}. The location of the event horizon is determined by the equation $A(r)=0$, so as seen in  Fig. \ref{Fig0}, this occurs in smaller distances in higher dimensions which asserts that for large distances gravity in four-dimensional spacetime is stronger than higher dimensions. This fact can also be checked by noting that, in higher dimensions, the $A(r)$ curves tend more rapidly to the $A(r)$ of flat spacetime.\\ The vierbein components of the vector $ u^{a}$  and $ \overrightarrow{B} $ are defined as follows \cite{l10,l11,l13}:

\begin{equation}\label{14}
u^{a}=(\gamma A,U_{r}A^{-1},U_{\theta}r,U_{\phi}r\sin\theta), \overrightarrow{B}=(U_{\phi}\cos\theta,-U_{\phi}A\sin\theta,U_{\theta}A)
\end{equation}

To derive Eq.(\ref{14}) we used the fact that $ \overrightarrow{U}=\gamma \overrightarrow{v} $. Let us now  study  a neutrino orbiting a black hole. For simplicity, we consider only circular orbits with constant radius $ r $ $(U_{r}=\frac{dr}{d\tau}=0)$. It is important to note that not all the orbits with arbitrary radius are stable. Because of the spherical symmetry of the gravitational field, we constraint ourselves to the study of the orbits lying only in the equatorial plane $(\theta=\frac{\pi}{2}$ or equivalently $ v_{\theta}=0 $ so $ U_{\theta}=0 $ ).  $\Omega_{1}$ and $\Omega_{3}$ depend on $\cos\theta$ and $v_{\theta}$ so that in this case $\Omega_{1}= \Omega_{3}=0 $ \cite{l10,l11}. It should be noted that the vierbein four velocity now takes the subsequent form:
\begin{equation}\label{15}
u^{a}=(\gamma A,0,0,\gamma v_{\varphi}r).
\end{equation}

We also have the following forms for the gravi-electric and gravi-magnetic fields:

\begin{equation}\label{16}
\overrightarrow{E}=(E_{0},0,0),  \overrightarrow{B}=(0,-v_{\varphi} A \gamma,0),
\end{equation}

where $ E_{0}=\frac{1}{A^{2}(r)}(1-\frac{rF(r)}{2A^{2}(r)}) $, $v_{\varphi}$ and $\gamma^{-1}$ are defined a long these lines\cite{l14}:
\begin{equation}\label{17}
v_{\varphi}=\frac{d\varphi}{dt}=\sqrt{\frac{F(r)}{2r}},
\end{equation}

\begin{equation}\label{18}
\gamma^{-1}=\frac{d\tau}{dt}=\sqrt{A^{2}(r)-\frac{1}{2}rF(r)},
\end{equation}

where $F(r)=\frac{dA^{2}(r)}{dr}$.\\

After substituting Eq. (\ref{13}) in Eqs. (\ref{17}) and (\ref{18}), we have

\begin{equation}\label{19}
\gamma^{-1}=\sqrt{1-\frac{4L^{-3+d}\pi^{1-\frac{d}{2}}r^{2-d}\Gamma(\frac{d}{2})r_{g}}{-1+d}+\frac{2(2-d)L^{-3+d}\pi^{1-\frac{d}{2}}r^{2-d}\Gamma[\frac{d}{2}]r_{g}}{-1+d}},
\end{equation}
\begin{equation}\label{20}
v_{\varphi}=\sqrt{-\frac{2(2-d)L^{-3+d}\pi^{1-\frac{d}{2}}r^{-d}\Gamma[\frac{d}{2}]r_{g}}{-1+d}}.
\end{equation}

By substituting Eqs. (\ref{13}),(\ref{15}),(\ref{16}),(\ref{19}) and (\ref{20}) in Eq.~(\ref{9}) and then $ \overrightarrow{\Omega}=\frac{\overrightarrow{G}}{\gamma}$, the only non-zero component of the frequency of  neutrino spin oscillations in higher dimensions in Schwarzschild metric i.e .$\Omega_{2}$  is obtained:

$$\Omega_{2}=(+\frac{L^{-3+d}\pi^{1-\frac{d}{2}}r^{-d}\Gamma[\frac{d}{2}]r_{g}}{-1+d}-\frac{dL^{-3+d}\pi^{1-\frac{d}{2}}r^{-d}\Gamma[\frac{d}{2}]r_{g}}{2(-1+d)}$$
\begin{equation}\label{21}
-\frac{2dL^{-6+2d}\pi^{2-d}r^{2-2d}\Gamma[\frac{d}{2}]^{2}r_{g}^{2}}{(-1+d)^{2}}+\frac{d^{2}L^{-6+2d}\pi^{2-d}r^{2-2d}\Gamma[\frac{d}{2}]^{2}r_{g}^{2}}{(-1+d)^{2}})^{\frac{1}{2}}=-\frac{v_{\varphi}}{2}\gamma^{-1}
\end{equation}

\begin{figure}[H]
\centering
\includegraphics[width=3.5in]{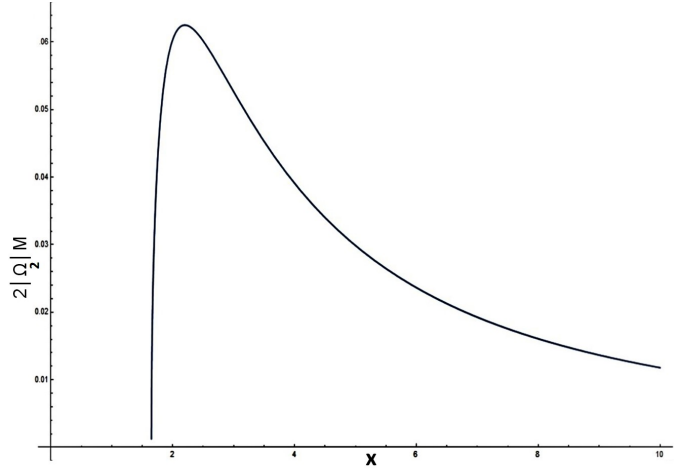}
\caption{\scriptsize Neutrino spin oscillations frequency for Schwarzschild metric vs. $ x $ for $ d=3 $ with $ \beta=2.4 $ and $ \alpha=1.5 $.\label{Fig1}}
\end{figure}

\begin{figure}[H]
\centering
\includegraphics[width=3.8in]{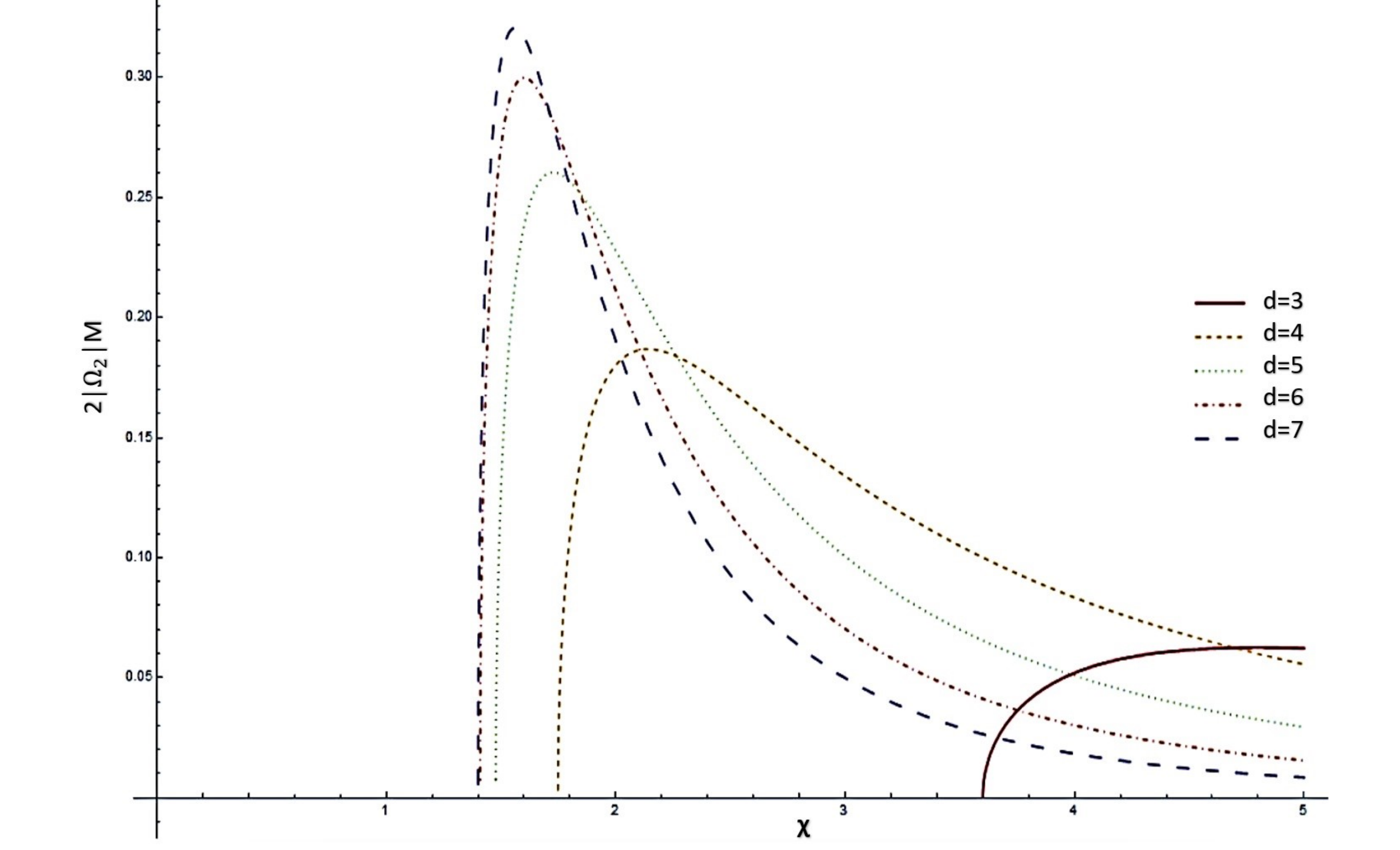}
\caption{\scriptsize (color online) Neutrino spin oscillations frequencies for Schwarzschild metric vs. $x$ for various dimensions with $ \beta=2.4 $ and $ \alpha=1.5 $. \label{Fig2}}
\end{figure}

We have plotted  $ 2|\Omega_{2}|M $ in Schwarzschild metric as a function of $ x $ for $ d=3 $ and $ d>3 $ in Figs. \ref{Fig1} and \ref{Fig2}.

It is seen from Fig. \ref{Fig1} that the neutrino spin oscillations frequency tends to zero at long distances.

 To draw Fig. \ref{Fig2}, we have set $ c=G=\hbar=1 $.  It is observed that the frequencies of oscillations decrease and tend to zero when $ x\rightarrow \infty $. It is also seen that the maximum of frequencies (peaks) for different dimensions have different values. By increasing the dimensions of space, the peaks occur in  smaller $ x $. Furthermore, it is noted that  for higher dimensions by increasing $x$, neutrino spin oscillations frequency decreases more rapidly than (3+1)-dimensional spacetime, which shows that for Large distances (compared to the size of extra dimensions) gravity becomes weaker in higher dimensions.\\

\begin{figure}[H]
\centering
  \includegraphics[width=3.2in]{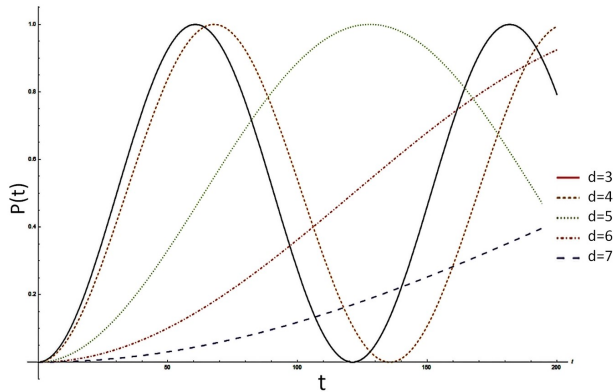}
\caption{\scriptsize (color online) Neutrino transition Probability for Schwarzschild metric. \label{Fig3}}
\end{figure}

Figure \ref{Fig3} shows the neutrino spin transition probability, $ P(t)=\sin^{2}(\Omega_{2}t) $. The period time of  oscillations increases as the dimension of space increases, which indicates that for short periods of time the contribution of three dimensional space to the probability of neutrino spin oscillation is more than higher dimensions.

\section{\large{Neutrino Spin oscillations  in  Reissner-Nordstrom (RN) Background in Higher Dimensions}}
RN metric describes the gravitational field of a charged non-rotating black hole, where its temporal component is defined as
\begin{equation}\label{22}
A(r)=\sqrt{1-\frac{r_{g}}{r}+\frac{Q^{2}}{r^{2}}}
\end{equation}

where $ Q $ is the charge of the black hole. The temporal component  in higher dimensions  takes the following form \cite{l16,l18}:

\begin{equation}\label{23}
A(r)=\sqrt{1-\frac{4r_{g}L^{d-3}\pi^{1-\frac{d}{2}}\Gamma[\frac{d}{2}]}{(d-1)r^{d-2}}+\frac{Q^{2}L^{2d-6}}{\frac{(d-1)(d-2)}{2}r^{2d-4}}}
\end{equation}

 By inserting Eq. (\ref{23}) in Eqs. (\ref{17}) and (\ref{18}) and doing  necessary calculations, the following expressions are obtained for the neutrino angular velocity and $ \gamma^{-1} $:

\begin{equation}\label{25}
  v_{\varphi}=\frac{d\phi}{dt}=( -\frac{2L^{-6+2d}Q^{2}r^{2-2d}}{-1+d}-\frac{2(2-d)L^{-3+d}\pi^{1-\frac{d}{2}}r^{-d}r_{g}\Gamma[\frac{d}{2}]}{-1+d} )^{\frac{1}{2}}.
\end{equation}

$$\gamma^{-1}=(1+\frac{2L^{-6+2d}Q^{2}r^{4-2d}}{(-2+d)(-1+d)}-\frac{4L^{-3+d}\pi^{1-\frac{d}{2}}r^{2-d}r_{g}\Gamma[\frac{d}{2}]}{-1+d}$$
\begin{equation}\label{24}
  +\frac{2L^{-6+2d}Q^{2}r^{4-2d}}{-1+d}+\frac{2(2-d)L^{-3+d}\pi^{1-\frac{d}{2}}r^{2-d}r_{g}\Gamma[\frac{d}{2}]}{-1+d})^{\frac{1}{2}}.
\end{equation}\\

\begin{figure}[H]
  \centering
  \includegraphics[width=3.2in]{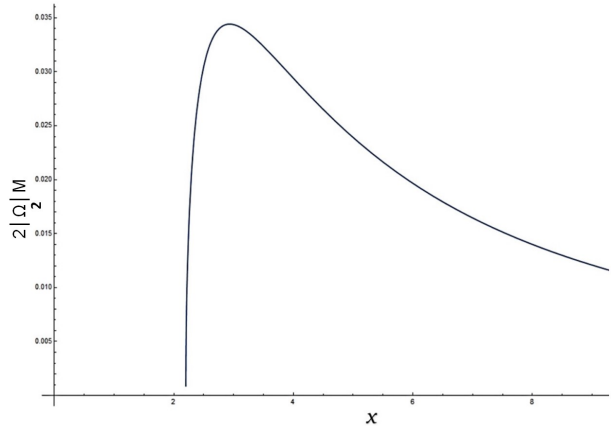}
\caption{\scriptsize Neutrino spin oscillations frequency for RN metric vs. $x$ in three dimension.\label{Fig4}}
\end{figure}

The only nonzero component of the  frequency in higher dimensions is obtained  by using Eqs. (\ref{23}),(\ref{25}),(\ref{24}),(\ref{15})and(\ref{16}), and inserting in relations (\ref{9}) and  $\overrightarrow{\Omega}=\frac{\overrightarrow{G}}{\gamma}$:

$$\Omega_{2}=(+\frac{L^{-12+4d}Q^{4}r^{6-4d}}{(-1+d)^{2}}+\frac{L^{-12+4d}Q^{4}r^{6-4d}}{(-1+d)^{2}(-2+d)}+\frac{L^{-6+2d}Q^{2}r^{2-2d}}{2(-1+d)}$$
$$+\frac{2L^{-9+3d}\pi^{1-\frac{d}{2}}Q^{2}r^{4-3d}r_{g}\Gamma[\frac{d}{2}]}{(-1+d)^{2}}+\frac{2L^{-9+3d}\pi^{1-\frac{d}{2}}Q^{2}r^{4-3d}r_{g}\Gamma[\frac{d}{2}]}{(-1+d)^{2}(-2+d)}$$
$$-\frac{2dL^{-9+3d}\pi^{1-\frac{d}{2}}Q^{2}r^{4-3d}r_{g}\Gamma[\frac{d}{2}]}{(-1+d)^{2}}-\frac{dL^{-9+3d}\pi^{1-\frac{d}{2}}Q^{2}r^{4-3d}r_{g}\Gamma[\frac{d}{2}]}{(-1+d)^{2}(-2+d)}$$
$$+\frac{L^{-3+d}\pi^{1-\frac{d}{2}}r^{-d}r_{g}\Gamma[\frac{d}{2}]}{-1+d}-\frac{dL^{-3+d}\pi^{1-\frac{d}{2}}r^{-d}r_{g}\Gamma[\frac{d}{2}]}{2(-1+d)}$$
$$-\frac{2dL^{-6+2d}\pi^{2-d}r^{2-2d}r_{g}^{2}\Gamma[\frac{d}{2}]^{2}}{(-1+d)^{2}}$$
\begin{equation}\label{26}
+\frac{d^{2}L^{-6+2d}\pi^{2-d}r^{2-2d}r_{g}^{2}\Gamma[\frac{d}{2}]^{2}}{(-1+d)^{2}})^{\frac{1}{2}}=-\gamma^{-1}\frac{v_{\varphi}}{2}.
\end{equation}
 To proceed further, we  define a new variable $\eta=\frac{Q}{\ell_{p}}$. In Fig.~\ref{Fig4}, we have plotted $ 2|\Omega_{2}|M $ in three dimensional RN metric as a function of  $ x $ for  $ \beta=2.4,  \eta=0.6 $ and $\alpha=1.5 $.

\begin{figure}[H]
  \centering
  \includegraphics[width=3.5in]{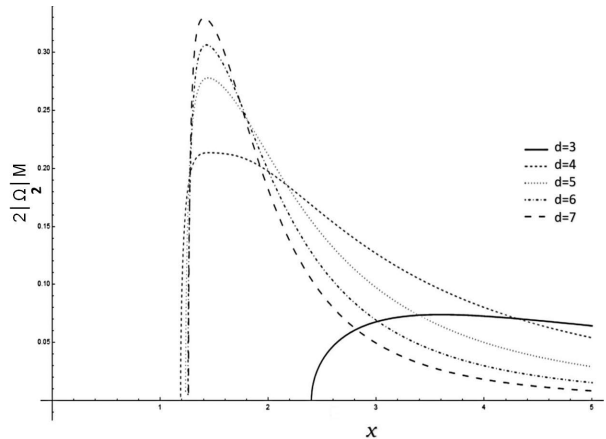}
\caption{\scriptsize  Neutrino spin oscillations frequency  vs. $ x $ for RN metric in higher dimensions. \label{Fig5}}
\end{figure}

 $ 2|\Omega_{2}|M $ in three as well as higher-dimensional RN metric as a function of $ x $ are presented in Fig.~\ref{Fig5}.

There are some  points that can be inferred from Fig. \ref{Fig5}: ( i ) There is a regular peak for the value of   frequency of oscillation in each spacetime dimension. ( ii ) The maximum values (peaks) reduce by decreasing the spacetime dimensions and the peaks of the frequencies take place in smaller x as  the number of dimensions  increase. ( iii ) As we get away from the horizon, the values of frequencies go faster to zero by increasing spacetime dimensions, which means that, in long distances, the dominant contribution belongs to three dimension, but is short distances (small $x$), the contributions of higher dimensions become stronger.
\begin{figure}[H]
  \centering
  \includegraphics[width=3.2in]{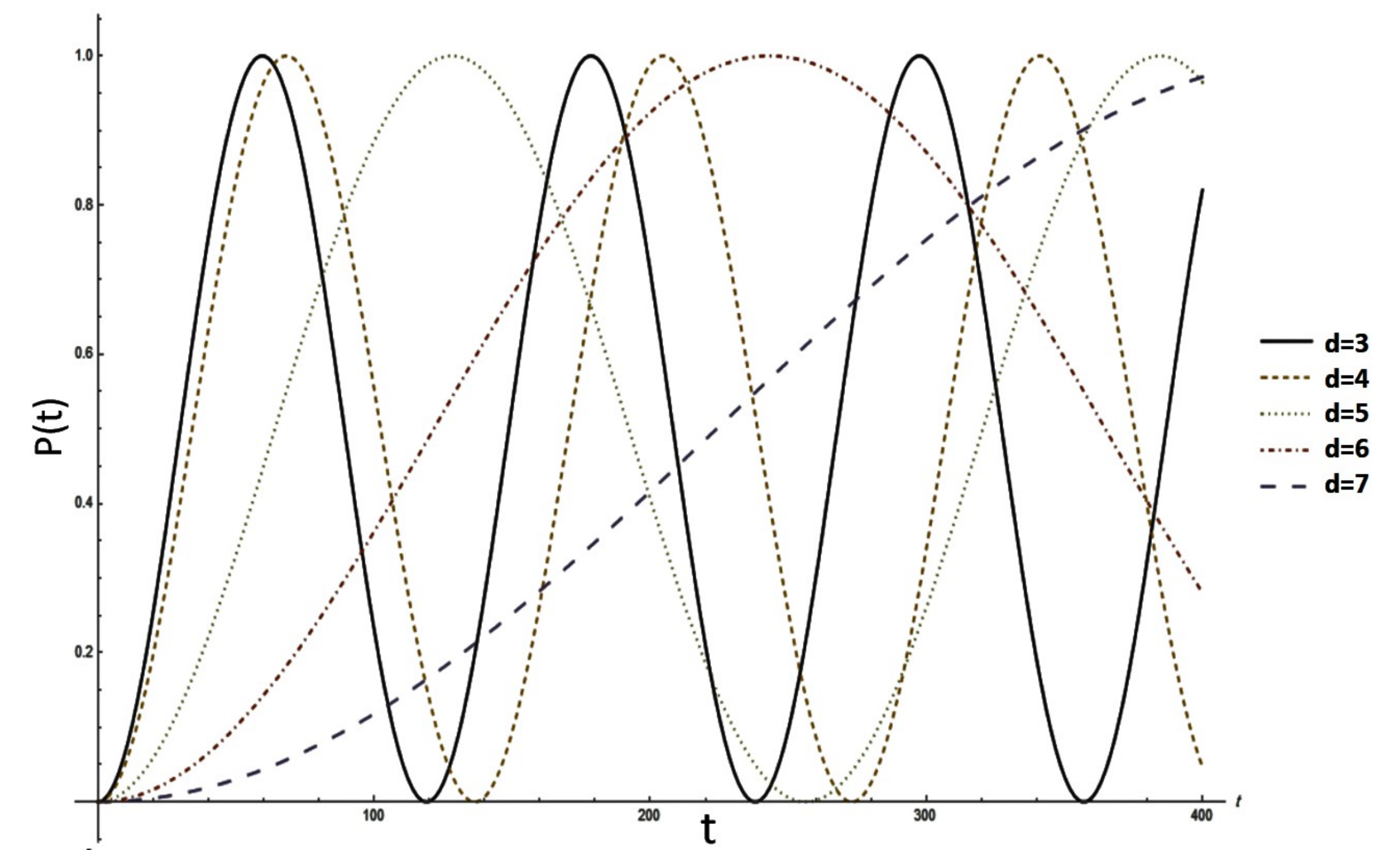}
\caption{\scriptsize Neutrino spin transition Probability for RN metric in Higher dimensions. \label{Fig6}}
\end{figure}
\begin{figure}[H]
  \centering
  \includegraphics[width=3.2in]{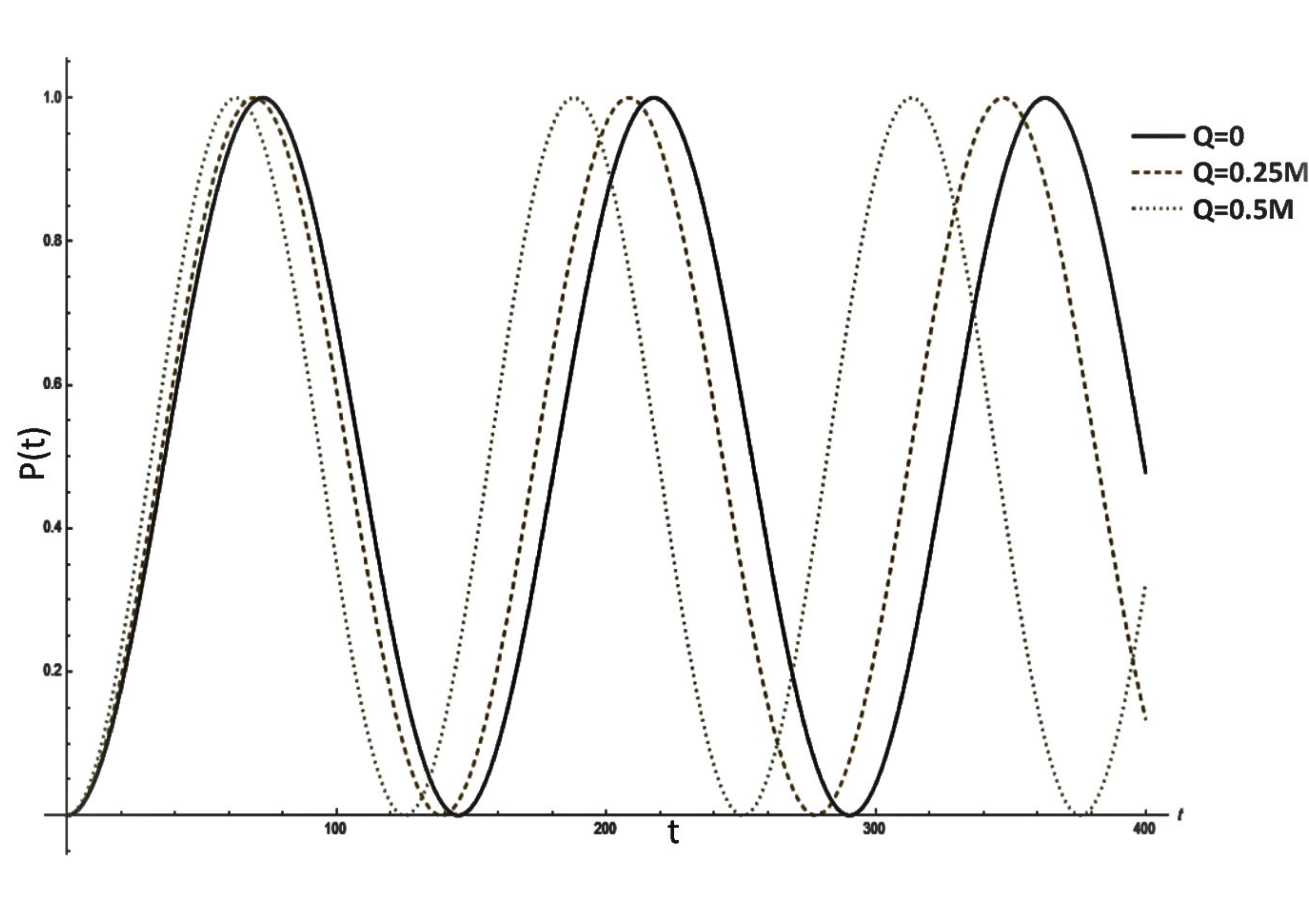}
\caption{\scriptsize Neutrino spin transition Probability for $Q=0, 0.25M, 0.5M$ in three dimension. \label{Fig7}}
\end{figure}

In  Fig.~\ref{Fig6}, the neutrino spin transition probability $ P(t) $ for RN metric is plotted  in three together with higher dimensions. It is seen that the behavior is the same as Schwarzschild metric i.e., by increasing the dimension of space, the time period of oscillations increases. Fig.~\ref{Fig7}  shows the effects of  charge of the black hole on neutrino spin transition probability $ P(t) $ in three dimensions. It is observed that by increasing the black hole charge the time period of oscillations decreases.\\\\

\begin{figure}[H]
\centering
  \includegraphics[width=5in]{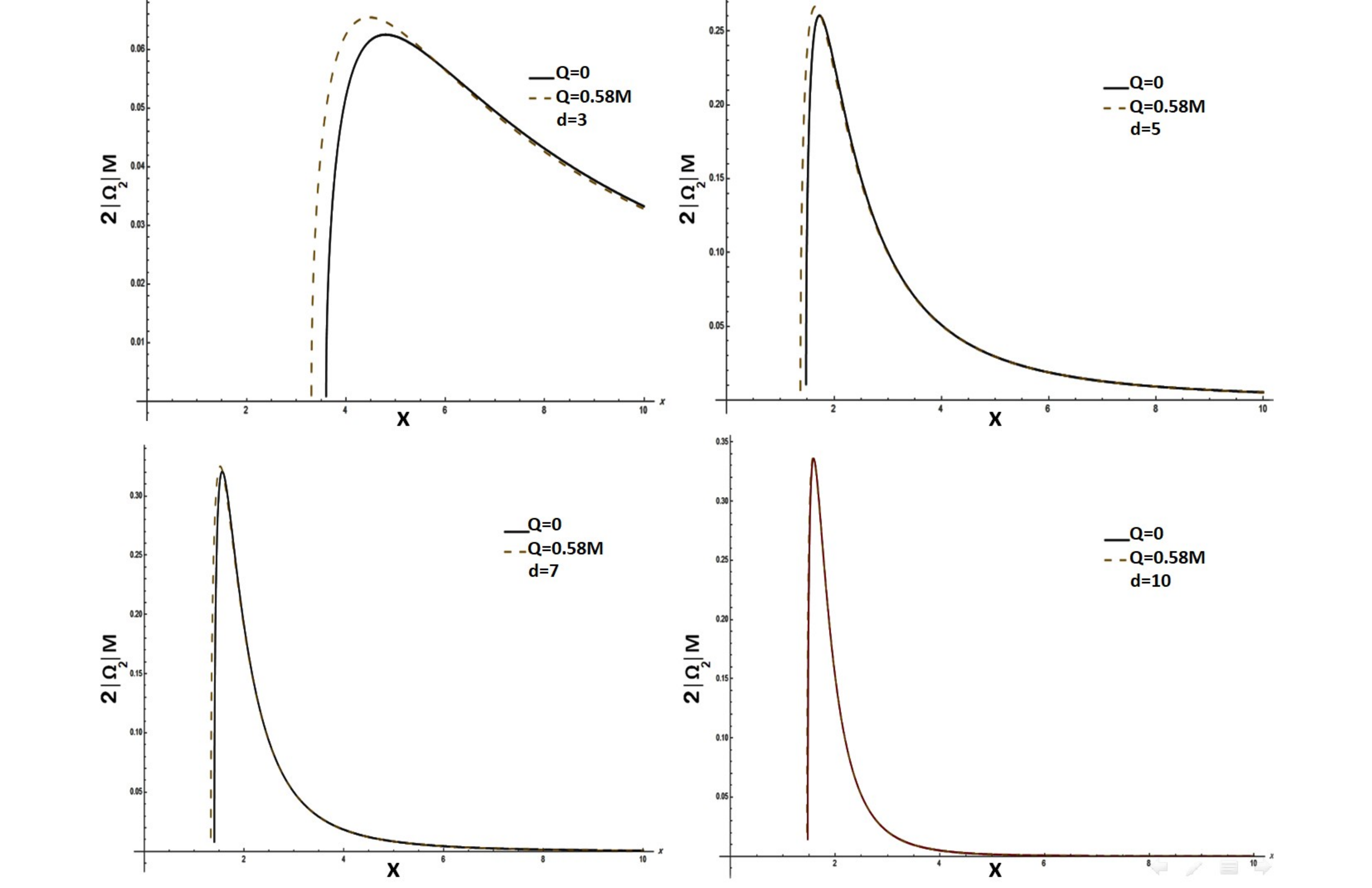}
\caption{\scriptsize The effects of the black hole charge on neutrino spin oscillation frequency for RN metric in different dimensions with $\eta=0.7$.\label{Fig10}}
\end{figure}
Figure \ref{Fig10} displays the effects of charge of the black hole on $\Omega_{2}$  in different dimensions. the following points can be deduced from the figure: (i) By increasing the dimension of space, the effects of the black hole charge decrease and diminish to zero. (ii) In each dimension by increasing the charge of the black hole, the peak occurs in smaller distances and finally the $Q=0$ curves are in agreement with the results obtained for the Schwarzschild metric.

\section{\large{Phenomenological Application}}

Let us  present a phenomenological application of our results. We consider a simple neutrino bipolar system which  helps us to understand many of the qualitative features of collective neutrino oscillations in supernovae. The system composed of a homogeneous and isotropic gas that initially consists of mono-energetic $ \nu_{e} $ and $ \overline{\nu_{e}}$ and is described by the flavor pendulum \cite{l20}. We introduce $ \epsilon $ as the fractional excess of neutrinos over antineutrinos, $ \epsilon=\frac{n_{\nu}}{n_{\overline{\nu}}} $. Now let us consider such systems in the presence of a neutral (Schwarzschild) black hole  with the same initial value of $ \epsilon $ in different dimensions:
\begin{equation}
\epsilon^{d=3}_{t=0}=\epsilon^{d=4}_{t=0}=.....\epsilon^{d=7}_{t=0}
\end{equation}

It can be checked from  Fig. \ref{Fig3} and \ref{Fig6} that the time period  of the neutrino  spin oscillations  probability increases with increasing dimensions:

\begin{equation}
T_{d=3}<T_{d=4}<T_{d=5}<T_{d=6}<T_{d=7},
\end{equation}

so at a later  time $ t>0 $, we have
\begin{equation}
\epsilon^{d=3}_{t}<\epsilon^{d=4}_{t}<\epsilon^{d=5}_{t}<\epsilon^{d=6}_{t}<\epsilon^{d=7}_{t}.
\end{equation}

The same result is obtained for the RN metric. This implies that the spin oscillations frequencies of the flavor pendulum as a function of the
neutrino number density for both  metrics decrease in higher dimensions.

\section{\large{Discussion}}

In this section, we discuss some more details and clarify some points. We start
with a discussion about the values of neutrino energy in the problem of spin oscillation around the black holes. There is a wide range of neutrino energies from
one-millionth of an electronvolt to $10^{18}$ electronvolt. This implies that there is a great deal of neutrinos to explore and getting interesting information about the processes that formed those neutrinos. Some of the lowest-energy neutrinos come from
the Big Bang (cosmic neutrino background or relic neutrinos), while the most energetic ones reach us from extragalactic sources. It is essential to recognize that the
cosmic neutrino background (CNB) is the main known source of non-relativistic
neutrinos in the universe. There have been much interest and efforts to detect
these non-relativistic neutrinos here on earth, see Ref. \citen{l21}. So in principle, there is
a broad possible range of energy for neutrinos but, it is worth mentioning that,
usually the spin-flip induced by only gravitation is strongly suppressed in the relativistic regimes and in fact, a magnetic field is needed.\cite{l22} A comment about the
phenomenological application is also in order. We have interpreted the neutrino
spin precession as neutrino-antineutrino oscillations, which is true for Majorana
neutrinos. In other words, for Majorana neutrinos, the neutrino spin flip is equivalent to neutrino-antineutrino oscillations. For the Dirac case, only the left-handed
neutrinos are available and the right-handed neutrinos’ population is not included
in the CNB or other neutrino energy ranges. Therefore, for the Majorana case,
the neutrino flux contains both left and right-helical neutrinos, and the capture
rate is increased. For Dirac neutrinos, after spin transition from left-handed to
right-handed neutrinos, the latter becomes sterile, and hence does not interact.
Consequently, the related flux should be modified by dimensional effects, along the
lines discussed by Dvornikov.\cite{l23,l24}

\section{\large{Conclusion}}

In this paper, we have investigated neutrino spin oscillations in gravitational fields in higher dimensions. We have analyzed the dependence of the oscillation frequency on the orbital radius. For both Schwarzschild and RN metrics, the maximum of the frequency increases by increasing the dimensions of space and the peaks occur in smaller distances compared to the three-dimensional case. It is also observed that for long distances the main contribution to the neutrino spin transition probability belongs to three dimensions. In addition, in both metrics, the period time of oscillation increases in extra dimensions. It is also shown that the effects of the black hole charge on the frequency of spin oscillations decrease and tend to zero in higher dimensions. We have also briefly studied the effects of higher dimensions on a bipolar neutrino system.\\\\

\end{document}